# Modulation of Field Emission Resonance on photodetachment of negative ions on surface

Yan Han, Lifei Wang, Ningyu Yang[*], Shiyong Ran and Guangcan Yang

School of Physics and Electronic Information, Wenzhou University,

Wenzhou,325035,China

## Abstract

The interaction between the field emission resonance states and the photodetached electron in an electric field is studied by semiclassical theory. An analytical expression of the photodetachment cross section is derived in the framework. It is found that the Stark shifted image state modulates the photodetachment cross section by adding irregular staircase or smooth oscillation in the spectrum. When the photodetached electron is trapped in Stark shifted image potential well, the detachment spectrum displays an irregular staircase structure which corresponds to the modified Rydberg series. While the photodetached electron is not bound by the surface potential well, the cross secton contains only a smooth oscillation due to the reflection of electronic wave by the field or the surface.

*Permanent address: Wenzhou No. 22 high school, Wenzhou, China.

Corresponding author: Guangcan Yang

Tel:86-577-86689033 Fax:86-577-86689010

Email:yanggc@wzu.edu.cn

# I. INTRODUCTION

Surface states have been studied intensively for over two decades in surface science. These states are trapped between the surface barrier potential and a band gap in the crystal. A typical example is an electron outside a metal surface weakly bound by the polarization it induces in the charge density in the near-surface region, which results in so-called image states that form a Rydberg-like energy level structure converging on the vacuum level[1-3]. The image states have been observed by inverse photoemission spectroscopy (IPES) [4-6] and two photo photoemission spectroscopy (2PPES)[3,7]. In the last dacade, time-resolved two-photon photoemission(TR2PPE) spectroscopy was used to study the electron dynamics of image potential states on metal surfaces [3,8]. And the results reveal the dynamical evolution of excited electrons in real time and the quantum beats spectroscopy are recorded to show the quantum coherent process[8,9]. On the other hand, the properties of atoms or ions adsorbed are influenced by the surface. For example, the negative ion resonance can be controlled by adsorption of the molecule onto an epitaxial metallic film [10]. And the presence of a surface alters the resonance lifetime of an adsorbed ion since the barrier penetrability of the detached electron is modified [11,12]. Actually, the existence of surface or interface modifies the spectral properties of atom and/or ion nearby[13,14]. It has been noted that the Rydberg-like series can be found in the photodetachment cross section of $H^-$ near a metal surface, which converges to the limit above the threshold, which is the emergence of image states formed by the photodetached electron near the metal surface[15].

In the mean while, scanning tunneling microscope (STM) can also be used to observe the image potential states[16,17]. In the case, the strong electric field near an STM tip creates a trapezoidal tunning barrier and distorts the series of bound image states[16, 18-20]. The states are Stark shifted to higher energy (often above the vacuum level of the unperturbed system), and the spacings between the states are expanded[16]. Therefore, these new features have considerable influence on the photo-atom/ion interaction near the surface. In the paper, we explore the photodetachment of $H^-$ near a metal surface in an electric field, which corresponds to the case when STM tip is applied to the surface.

We use semiclassical closed-orbit theory to investigate the photodetachment cross section of $H^-$ near the metal surface [21,22]. The closed-orbit theory emphasizes the role of closed orbits and their repetitions. The closed orbits are those classical trajectories which go out from the origin where the nucleus or atomic core is located and turned back by external fields and/or other potential barriers to the vicinity of origin. In the semiclassical framework, electronic waves travel along the closed orbits and interfere with the outgoing wave source. We obtain an analytical expression for the cross section as the sum of the contributions of the closed orbits and their repetitions.

The paper is organized as follows. In Sec. II we describe the classical motion of the photodetached electron in various configuration. All the closed orbits are obtained and the related classical actions, orbit periods, and Maslov index are calculated analytically. According to all the classical quantities, we derive an analytical formula of the photodetachment cross section of the hydrogen ion H⁻ near metal surface in an electric field in Sec. III by closed-orbit theory. Some numerical results and conclusive remarks are presented in Sec. IV. Unless indicated otherwise, we use atomic units throughout this work.

## II. DYNAMICS OF PHOTODETACHED ELECTRON

The system configuration is set up as follows: In cylindrical coordinates ($\rho, z, \phi$), a hydrogen negative ion H⁻ locates at the origin and a $z$-polarized laser is applied for the photodetachment. A metal surface perpendicular to z-axis is put at $-z_0$ where $z_0$ is positive, and an electric field in $z$ direction is applied. In the measurement of surface states by scanning tunning microscope (STM) such electric field exists in the vacuum gap between the tip and sample. The Coulomb interaction between the photodetached electron and its positive image charge formed by the polarization of the nearby conduction band electrons take the form $V(z) = -\frac{1}{4(z_0 + z)}$. In this configuration, the Hamiltonian governing the photodetached electron motion can be written

$$H = \frac{p_\rho^2 + p_z^2}{2} + V_b(r) - \frac{1}{4(z + z_0)} + F(z + z_0), \qquad (1)$$

where $F$ is the strength of electric field. In Eq.(1), $V_b(r)$ is the short-range potential of the polarized central field of H atom, can be ignored in the calculation of classical motion of the photodetached electron far from the neutral core. Its quantum effect can be included in the initial outgoing electroic wave. The $\phi$ motion has been separated, and the z component of the angular momentum is a constant of motion which has been set to zero due to the cylindrical symmetry of the system. The Hamiltonian is separable and the classical motion of the electron can be separated into the motion along the *z* axis and the motion in *x-y* plane. The trajectories in the two direction can be formally written as

$$z(t) = \int_0^t \sqrt{2k^2 \cos^2\theta + \frac{1}{2(z+z_0)} + \frac{1}{2z_0} - 2Fz}\, dt$$
$$\rho(t) = kt\sin\theta$$

(2)

where *k* is the initial momentum.

For simplicity we let $d = z_0 + z$, i.e. the distance between detached electron and metal surface. Thus the surface potential can be rewritten as $V(d) = -\frac{1}{4d} + Fd$. Figure 1 shows the schematic surface potential for different electric field configurations and the location of negative ion. The electronic dynamics can be diescribed as follows:

(1) For the case of $F > 0$, the electric field shifts up the image potential as shown in Fig. 1(a), the photodetached electron is always trapped in the surface quantum well regardless the location of the negative ion. The electron moves freely in $\rho$ plane, and oscillates back and forth in *z*-direction. In all the classical trajectories of the photodetached electron emanating out the origin, those bounced back by Stark shifted image potential to the starting point are called closed orbits. The following are fundamental. (i) The electron goes up along the *+z* direction, reaches its maximum and then bounced back by the potential and returns to the origin.

We call this orbit the up orbit. (ii) The electron goes down in –z direction and hits the metal surface and bounces back and finally return to the origin. We call this orbit the down orbit. (iii) The electron completes the up orbit first and then passes through the origin, and continues to complete the down orbit. (iv) The electron completes the down orbit first and then up orbit. This orbit is similar to the one of (iii) but in reverse order.

(2) For the case of $F < 0$, the electric field pulls down the image potential as is shown in Fig.1(b). When $d = d_0 = \sqrt{-\dfrac{1}{4F}}$, potential V(d) arrives at its maximum value $V_{max} = -\sqrt{-F}$. Thus a potential barrier occurs at $d_0$. The motion of photodetached electron varies with the distance $z_0$ between the negative ion H⁻ and mental surface. When $z_0 < d_0$, there are three different motions according to the energy of the photodetached electron.

  (a) The energy of photodetached electron is lower than the barrier. The electron is still trapped in the surface potential well and its classical motion is similar to the case of (1).

  (b) The energy of photodetached electron is equal to the barrier. If the electron goes up along the +z direction, it would reaches potential peak at which its kinetic energy vanishes and would stop keeping an unstable balance. The electron that goes down along the -z direction will hits the metal surface and be bounced back and finally return to the origin then goes up along the +z direction as is stated above. So only down orbit without repetition exists in this case.

(c) The energy of photodetached electron is higher than the barrier. The electron that goes up along the +z direction will escape from the bind of the potential without returning. Only down orbit is closed orbit and no orbit repetition exists in this case.

(3) For the case of $F < 0$ and $z_0 = d_0$, as long as the electron has a nonzero kinetic energy, the electron that goes up along the +z direction will escape from the bind of the potential without returning. The electron that goes down along the -z direction will hits the metal surface and be bounced back and finally return to the origin then goes up along the +z direction escaping from the bind of the potential. Only down orbit without repetition exists in this case.

(4) For the case of $F < 0$ and $z_0 > d_0$, whatever the energy of detached electron is, the electron in +z direction escape the surface acceleratedly. But for the electron in –z direction, there are three different motions according to the energy of the photodetached electron.

(a) The energy of photodetached electron is lower than the barrier. The detached electron will reach its highest before it reaches the barrier and then be pulled back by the electric field. After it returns to origin it will go along +z direction then escape the surface. Only down orbit is closed orbit and no orbit repetition exists in this case.

(b) The energy of photodetached electron is equal to the barrier. The

detached electron will reaches potential peak at which its kinetic energy vanishes and will stop there keeping an unstable balance. So none closed orbit exist in this case.

(c) The energy of photodetached electron is higher than the barrier. the electron goes uphill in -z direction and crosses the potential peak, then enter the surface potential well. It experiences an oscillation in the well, finally returns the starting point. So only down orbit is closed orbit and no orbit repetition exists in this case.

Therefore, for $F<0$ and $z_0 > d_0$, the orbit has a transition when the energy increases gradually. In the case we have one closed orbit although the higher energy orbit is quite complicated.

For the cases (1) and (2a), we can classify the system of orbits as follows: indices $j$ and $n$ are used to label the closed orbits, where $j=1,2,3,4$ and $n=0,1,2,3,\cdots$. Here $n=0$ means that the orbit is a fundamental closed orbit ($j=1,2,3,4$ for the fundamental closed orbits above, respectively). When $n>0$, the orbit ($j,n$) has two parts, the beginning part and the later part. The beginning part is always the $j$th fundamental closed orbit. The later part consists of $n$ repetitions of periodic orbits $j=3$ or 4. For the case of $j=1$, it is $n$ repetitions of $j=3$, and for $j=2$, it corresponds to $n$ repetitions of $j=4$. For the periodic orbit $j=3$ or $j=4$, it repeats itself $n$ times. The classical quantities needed for constructing semiclassical wave function can be obtained according to the systematic procedure of closed orbit theory[21,22]. In the present case, if the initial kinetic energy of photodetached electron is $E$, we have $E = \dfrac{p_z^2}{2} - \dfrac{1}{4(z_0+z)} + \dfrac{1}{4z_0} + Fz$ where $p_z$ is the momentum in $z$-direction. The traveling times for classical orbits can be obtained by the integral identity $t = \int_0^t dt = \int_0^z \frac{dz}{p_z}$ of $z$-motion. For the fundamental closed orbits, their returning time $T_j$ can be written as

$$T_1 = \frac{4\sqrt{2}}{\sqrt{F}} \frac{1}{C} \left[ E(A_2 | B) - E(A_1 | B) \right]$$
$$T_2 = \frac{4\sqrt{2}}{\sqrt{F}} \frac{1}{C} \left[ E(A_4 | B) - E(A_3 | B) \right] \quad (3)$$
$$T_3 = T_1 + T_2$$
$$T_4 = T_1 + T_2$$

where E(A|B) is the incomplete elliptic integral of the second kind[23], and

$$A_1 = \frac{\sqrt{2}}{2} \sqrt{\frac{z_0 - b + \sqrt{a}}{\sqrt{a}}}$$
$$A_2 = \frac{\sqrt{2}}{2} \sqrt{\frac{z_{max} + z_0 - b + \sqrt{a}}{\sqrt{a}}}$$
$$A_3 = \frac{\sqrt{2}}{2} \sqrt{\frac{-b + \sqrt{a}}{\sqrt{a}}} \text{ or } A_3 = \frac{\sqrt{2}}{2} \sqrt{\frac{z_{min} + z_0 - b + \sqrt{a}}{\sqrt{a}}} \text{ (for case (4a))} \quad (4)$$
$$A_4 = \frac{\sqrt{2}}{2} \sqrt{\frac{z_0 - b + \sqrt{a}}{\sqrt{a}}}$$

$$B = \sqrt{-\frac{2\sqrt{a}}{-\sqrt{a} + b}} \quad (5)$$

$$C = \frac{2}{\sqrt{-\sqrt{a} + b}} \quad (6)$$

with

$$a = \left(\frac{E}{2F} - \frac{1}{8Fz_0} + \frac{z_0}{2}\right)^2 + \frac{1}{4F}$$

$$b = \frac{E}{2F} - \frac{1}{8Fz_0} + \frac{z_0}{2}$$

$$z_{max} = \frac{-(1+4z_0^2 F - 4z_0 E) + \sqrt{(1+4z_0^2 F - 4z_0 E)^2 + 64z_0^3 FE}}{8z_0 F} \quad (7)$$

$$z_{min} = \frac{-(1+4z_0^2 F - 4z_0 E) - \sqrt{(1+4z_0^2 F - 4z_0 E)^2 + 64z_0^3 FE}}{8z_0 F}$$

Clearly, we have $T_3 = T_4 \equiv T$. The classical action of a trajectory is defined as $S(\rho, z, \phi) = \int_0^t \mathbf{p} d\mathbf{q}$, where $t$ is the traveling time, $\mathbf{p}$ and $\mathbf{q}$ are coordinate and momentum vectors respectively. It is accumulated along a classical trajectory and appears in the phase of the semiclassical wave function. For the fundamental closed orbits, the returning times in Eq.(1) are used in the calculation of the classical actions. The classical actions along the fundamental closed orbits are

$$S_1 = -\frac{4\sqrt{2F}}{3}\left\{D_2 - D_1 - C(a - b^2)\left[F(A_2 | B) - F(A_1 | B)\right]\right.$$
$$\left. + Cb(\sqrt{a} - b)\left[E(A_2 | B) - E(A_1 | B)\right]\right\}$$

$$S_2 = -\frac{4\sqrt{2F}}{3}\left\{D_4 - D_3 - C(a - b^2)\left[F(A_4 | B) - F(A_3 | B)\right]\right. \quad (8)$$
$$\left. + Cb(\sqrt{a} - b)\left[E(A_4 | B) - E(A_3 | B)\right]\right\}$$

$$S_3 = S_1 + S_2$$
$$S_4 = S_1 + S_2$$

where F(A|B) is the incomplete elliptic integral of the first kind[23], E(A|B) is the incomplete elliptic integral of the second kind, and $A_1$, $A_2$, $A_3$, $A_4$, B, C, $a$, $b$, $z_{max}$, $z_{min}$ are defined in equation(4), (5), (6), (7) and

$$D_1 = -\sqrt{a-(z_0-b)^2}\sqrt{z_0}$$
$$D_2 = 0$$
$$D_3 = 0 \quad (9)$$
$$D_4 = -\sqrt{a-(z_0-b)^2}\sqrt{z_0}$$

Clearly, we have $S_3 = S_4 \equiv S$. Maslov index is related with the topological structure of trjectory manifold and can be obtained simply by counting the singular points such as caustics and foci along the trajectory[24]. For the current case, Maslov index can be easily found by counting the returning points where a $\pi$ phase loss of electronic wave occurs, thus we have

$$\mu_1 = 1, \mu_2 = 2,$$
$$\mu_3 = \mu_4 = 3. \quad (10)$$

The action of any closed orbit consists of the ones of fundamental orbits and can be written as

$$S_{jn} = S_j + nS. \quad (11)$$

The returning time of any closed orbit is

$$T_{jn} = T_j + nT \quad (12)$$

with Maslov index

$$\mu_{jn} = \mu_j + 3n. \quad (13)$$

For the case (2b), (2c), (3), (4c), only down orbit is closed orbit and no orbit repetition exists. $T_2$ and $S_2$ take the same form as is described in formulae (3), (4), (5), (6), (7), (8), (9).

For the case (4a), only down orbit is closed orbit and no orbit repetition exists. But this closed orbit is different from those in case (2b), (2c), (3), (4c) because the detached electron doesn't arrive at metal surface but at a lowest point and then be pulled back by electric field. In this case $T_2$ and $S_2$ take the same form as in case (1) except $A_3$ and $D_3$ having different forms which have been expressed in formulae (4) and (9).

For the case (4b), there is no closed orbit. So the parameters stated above don't exist.

## III. PHOTODETACHMENT CROSS SECTION

The photodetachement of $H^-$ in external fields or quantum well can be regarded as a one-electron process if we neglect the influence of the short-range potential $V_b(r)$ when the electron is far from the core. The electronic binding energy $E_b = k_b^2/2$ is approximately 0.754 eV, where $k_b$ is related with the initial wave function $\Psi_i = M \exp(-k_b r)/r$, where $M$ is a "normalization function" constant and is equal to 0.31552. When an incident laser beam is applied to the ion, the valence electron absorbs a photon energy $E_p = E_b + E$ and is photodetached. In Green's function formalism of optical absorption theory of atom or ion, the photodetachment cross section can be expressed in the form[25]

$$\sigma(E) = -\frac{4E_p}{c} \operatorname{Im} \left\langle D\Psi_i \middle| \hat{G}^+ \middle| D\Psi_i \right\rangle, \tag{14}$$

where $\hat{G}^+$ is the outgoing Green's function and is related with the external potentials. The dipole operator $D$ is equal to the projection of the electron coordinate onto the direction of polarization of the laser field. In Eq(14), the initial state is modified by the dipole operator related with the incident laser field to become the source wave function; the Green's function propagates these waves outward to become the outgoing waves; and finally the waves overlap with the source wave to give the absorption spectrum. The key point is how to obtain the outgoing Green's function. In principle we can compute the function by solving the inhomogeneous Schrödinger equation with source $D\Psi_i$. But the task usually is difficult and cumbersome both analytically and numerically for most real systems. An alternative is semiclassical method, whose advantage is independent of dimensionality and its clear dynamical picture. Closed orbit theory is developed by propagating the outgoing electronic wave semiclassically to solve the spectral problems of atoms or ions in external fields or quantum wells [13,15,21,22].

In closed orbit theory, we have the following picture: Near the atomic core, the electron moves in a straight line at constant speed, which can be described by the steady outgoing electronic spherical wave outward in all directions. At some distance from the core, the electronic wave feels the external fields or potentials and it propagates continually but distorted by the external fields or potentials. The propagation is achieved semiclassically by relating the wavefunction with classical trajectories. Finally, some of the waves pull back to the vicinity of the core and interfere with the outgoing wave to form the spectral pattern of the photodetachment. The outgoing wave can be divided the direct part and returning part, the former never goes far from the core and the latter propagates outward into the external region first, then is reflected by the potential or metal surface, and finally returns to the vicinity of the core to interfere with the outgoing wave. That is $\left|\hat{G}^+\right|D\Psi_i\rangle = \Psi_{dir} + \Psi_{ret}$. The cross section has two parts, $\sigma(E) = \sigma_0(E) + \sigma_{ret}(E)$, where the first part is the contribution of the direct wave interfering with the source and is the field-free cross section $\sigma_0(E) = \frac{16\sqrt{2}M^2\pi^2}{3c}\frac{E^{3/2}}{(E_b+E)^3}$ [24], and the second part is the contribution of the returning wave,

$$\sigma_{ret}(E) = -\frac{4E_p}{c}\text{Im}\langle D\Psi_i | \Psi_{ret}\rangle, \qquad (16)$$

where $\Psi_{ret}$ is the returning wave, which overlaps with the outgoing source wave to give the interference pattern in the absorption spectrum.

We construct the returning wave $\Psi_{ret}$ semiclassically, then calculate the overlapping integral in Eq.(16) [25,27]. The procedure can be presented briefly as follows:

When a beam of $z$-polarized light is applied to the initial state of electron, the outgoing electron wave is produced. On the surface of the virtual sphere, the outgoing wave can be written as[28]

$$\Psi^0(R,\theta,\phi) = -i\frac{4Mk^2}{\left(k_b^2+k^2\right)^2}\cos(\theta)\frac{e^{i(kR-\pi)}}{kR}. \qquad (17)$$

where R is the radius of the sphere, which is arbitary and the final result is independent of its value. A reasonable choice is about $R \approx 10a_0$. $\Psi^0(R,\theta,\phi)$ is the initial wave function for propagating outward. In the semiclassical approximation, the wave outside this sphere can be expressed as

$$\Psi(\rho,z,\phi) = \sum_i \Psi^0(R,\theta,\phi)A_i e^{i[S_i - \mu_i\pi/2]} \qquad (18)$$

where $S_i$ is the action along the $i$th trajectory, $\mu_i$ is the Maslov index characterizing the geometrical properties of the $i$th trajectory and its neighboring orbits. The amplitude of the wave function $A_i$, is related with the divergence of adjacent trajectories from a central trajectory, is given by[15]

$$A_i(\rho,z,\phi) = \left|\frac{J_i(\rho,z,0)}{J_i(\rho,z,T)}\right|^{1/2} = \frac{R}{R+kT}. \qquad (19)$$

where Jacobian is defined as

$$J(\rho,z,\phi) = \rho(T)\begin{vmatrix} \frac{\partial z}{\partial t} & \frac{\partial z}{\partial \theta} \\ \frac{\partial \rho}{\partial t} & \frac{\partial \rho}{\partial \theta} \end{vmatrix}. \qquad (20)$$

In eqs.(19) and (20), $T$ represents the traveling time of the photodetached electron from the origin. The returning waves are associated with the closed orbits returning to the vicinity of the core. And the returning waves near the core must be cylindrically symmetric, and can be approximated by incoming Bessel functions. The returning wave function related with closed orbit $(j,n)$ must match the Bessel function and can be written as

$$\Psi_{jn}^{ret} = N_{jn}\frac{1}{\sqrt{2\pi}}J_0(k_\rho^{ret}\rho)\frac{1}{\sqrt{2\pi}}e^{ik_z^{ret}z}, \qquad (21)$$

where the normalization factor $N_i$ can be determined by matching Eq.(18) and Eq.(21). For closed orbit $(j,n)$, the modified amplitude in Eq.(13) of the returning function is

$$A_{jn} = \frac{1}{T_{jn}k}. \qquad (22)$$

The proportional constant in Eq.(15) can be found by matching Eq.(21) to Eq.(18) and can be written as

$$N_{jn} = (-1)^{[(\mu_j-1)/2]}\frac{4iMk^2}{T_{jn}k\left(k_b^2+k^2\right)^2}\exp\left[\frac{i}{\hbar}\left(S_{jn}-\mu_{jn}\frac{\pi}{2}\right)\right] \qquad (23)$$

The returning part of the cross section can now be calculated by overlapping the returning waves with the outgoing wave functions, then taking the imaginary part[13,25]. Therefore, we have

$$\sigma_{ret}(E) = -\frac{4E_p}{c}\text{Im}\int M\exp(-k_b r)\cos\theta N_{jn}e^{-ikr\cos\theta_{jn}^{ret}}r^2 dr\sin\theta d\theta d\phi, \quad (24)$$

Integrating out the expression and adding the field free part, we have the total photodetachment cross section

$$\sigma(E) = \frac{16\sqrt{2}M^2\pi^2}{3c}\frac{E^{3/2}}{(E_b+E)^3} - \sum_{jn}C_{jn}\sin(S_{jn}-\mu_{jn}\pi/2). \quad (25)$$

where c is light speed and

$$C_{jn} = (-1)^{[(\mu_j-1)/2]}\frac{2\pi^2}{c}\frac{4\sqrt{2}M^2 E^{1/2}}{T_{jn}(E_b+E)^3}. \quad (26)$$

It is noted that Eq.(25) is the same as in the case of without electric field[15], but now we have different classical action and returning time. When $z$ motion and $\rho$ motion are seperable, Eq.(25) is generally valid for surface photodetachment.

## IV. NUMERICAL RESULTS AND CONCLUSIONS

The photodetachment cross section can be calculated for different configurations described in Sec. II. The calculated photodetachment spectra for the case of positive electric field are shown in Fig. 2 to 4. In the case, a surface quantum well is formed between the metal surface and the Stark lifted image potential. Thus, the photodetached electron is trapped in the well because of the infinite well depth. As discussed in Sec. II, there are four fundamental closed orbits and their repetitions. The returning waves from the closed orbits interfere with the steadily outgoing waves constructively or destructively to shape up the spectral structure of photodetachment. In Fig 2 to 4, the staircase structure results from the superposition of contribution from the closed orbits in the quantum well. The electronic waves are bounced back and forth between the metal surface and Stark lifted image potential, the spectrum displays a staircase structure, which comes from the quantized energy level of the well actually. Fig. 2 shows the photodetachment cross of the negative ion in a weak electric field. When the energy is slightly higher than the threshold, we can see an explicit staircase structure like Rydberg series which corresponds to the weakly shifted image states. When the energy goes up, the cross section show an almost uniform sawtooth structure which corresponds to the quite uniform energy level interval in a triangular potential obtained by dropping the image potential in Eq.(1). In the Hamiltonian of Eq.(1) by neglecting the image potential, the eigen energy level $E_n = \left(\frac{F^2}{2}\right)^{1/3} \left[\frac{3\pi}{2}\left(n-\frac{1}{4}\right)\right]^{2/3} - Fz_0$, and the space of adjacent levels for large $n$ is almost uniform. We show the case of medium electric field in Fig. 3, where the image energy level structure can be seen on the first or second sawtooth near the threshold while the remaining is basically the approximate uniform level structure due to the triangular potential. In Fig. 4, we can see that the intervals between the neighboring sawteeth increase with the electric field strengthes. The case of negative electric field is shown in Fig. 5 and 6. In Fig. 5, the negative ion is located in the surface well with a sallow barrier. For the case, the photodetached electron oscillates in the well when the energy is lower than the barrier and a staircase structure forms in the photodetachment spectra but with very few steps since there are quite few bound states in the sallow potential well. When the energy is higher than the barrier, the photodetached electrons in +z direction overcomes the barrier and escapes from the well directly while those in –z direction accelerate first, are reflected by the metal surface, pass the starting point and finally escape from the well. In the case, there is only one down closed orbit, which leads to an oscillation in the higher energy part of the photodetachment cross section. When the ion is outside the barrier, we have a different scenario with the cases discussed. In the case, the only closed orbit is in –z direction. When the electron has lower energy, it climbs up the potential hill to some height and is pulled back by the electric field to form a closed orbit. When the energy is higher than the potential barrier, the electron passes the barrier and enters the surface potential well, bounces back by the metal surface, returns to the origin. Clearly, these two energy ranges show very different classical actions and amplitudes in Eq.(25). We expect a transition at the barrier energy point. It is confirmed in the photodetachment spectra and the transition is shown in Fig. 6.

In the present investigation, we didn't consider the influence of lifetimes of the image states and assumed that they are infinite long. This is certainly not true for real systems. In fact, the lifetimes of image states and Starked shifted image states depend on the type of metals, crystal orientation and the atom adoption of surface [20, 29, 30]. For a finite lifetime system, the contribution from those orbits whose periods are longer than the lifetime vanishes in Eq.(19), but the orbits of short periods still contribute to the summation. In the case, there still exists the staircase structure in the cross section but with smoother steps due to the lack of high harmonics unless the shortest period of closed orbits is longer than the lifetime. If there is further dephasing such as surface atom adsorption, the staircase structure may disappear in the cross section. Further investigation is needed to clarify the subject in detail.

In summary, the interaction between the field emission resonance states and the photodetached electron in an electric field has been studied by semiclassical theory. It is found that the Stark shifted image state modulates the photodetachment cross section by adding irregular staircase or smooth oscillation in the spectrum. When the photodetached electron is trapped in Stark shifted image potential well, the detachment spectrum displays an irregular staircase structure which corresponds to the modified Rydberg series. While the photodetached electron is not bound by the surface potential well, the cross section contains only a smooth oscillation due to the reflection of electronic wave by the field or the surface.

**Acknowledgement**

This work is partially supported by a National Key Basic Research Project of China (2007CB310405).

# FIGURE CAPTIONS

Figure 1: The surface potential for different electric field configurations. (a) $F > 0$, the electric field shifts up the image potential. (b) $F < 0$, the electric field pulls down the image potential and a potential barrier occurs at $d_0$.

Figure 2: The photodetachment cross sections of the negative ion located at two different positions with values of $z_0$ 50 a.u. and 100 a.u. respectively in a weak electric field F=50kV/cm. When the energy is slightly higher than the threshold, we can see the explicit staircase structure like Rydberg series which corresponds to the weakly shifted image states. When the energy goes up, the cross section shown a almost uniform sawtooth structure which corresponds to the quite uniform energy level space in a triangular potential.

Figure 3: Medium electric field F=500kV/cm corresponding to Fig. 2. The image energy level structure can be seen on the first or second sawtooth near the threshold while the remaining is basically the approximate uniform level structure due to the triangular potential.

Figure 4: Comparison between two different cross section under electric field F=50kV/cm and F=200kV/cm respectively. Where we set $z_0 = 100 a.u.$. we can see that the intervals between the neighboring sawteeth increase with the electric field strengthes.

Figure 5: The negative ion is located in the surface well with a sallow barrier. Here $d_0 =161$ a.u. corresponding to the electric field $F = -50 kV/cm$ we take. When the energy is lower than the barrier very few staircase structures forms in the photodetachment spectra which appears smooth oscillation when the energy is higher than the barrier.

Figure 6: the negative ion is outside the barrier. Here $d_0 = 51$ a.u. corresponding to the electric field $F = -500 \text{kV/cm}$. When the electron has lower energy, it climbs up the potential hill to some height and is pulled back by the electric field to form a closed orbit. When the energy is higher than the potential barrier, the electron passes the barrier and enters the surface potential well, bounces back by the metal surface, returns to the origin. A transition at the barrier energy point is found in the photodetachment spectra.

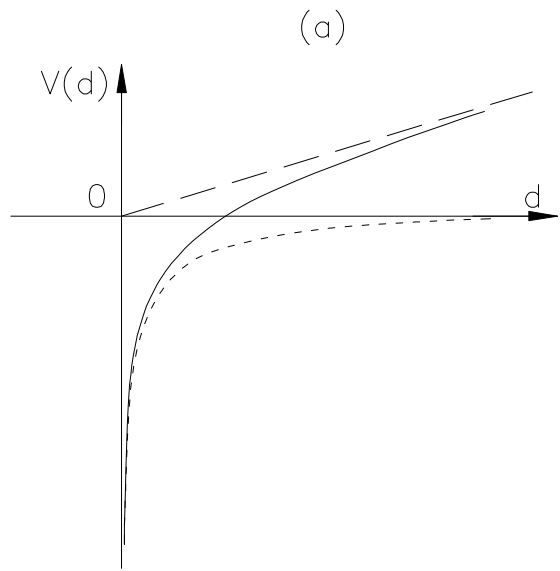

Fig1. (a)

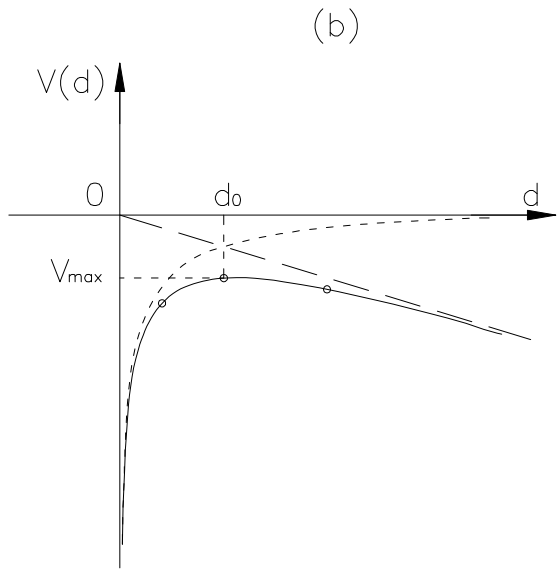

Fig1. (b)

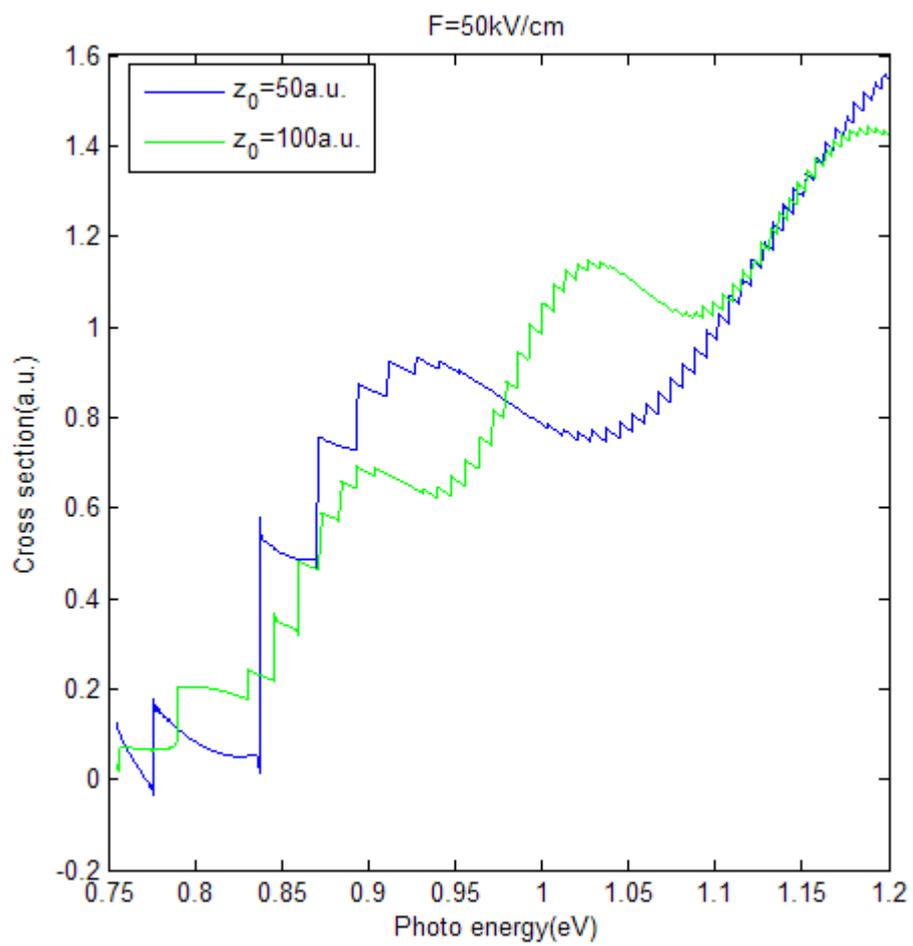

Fig.2

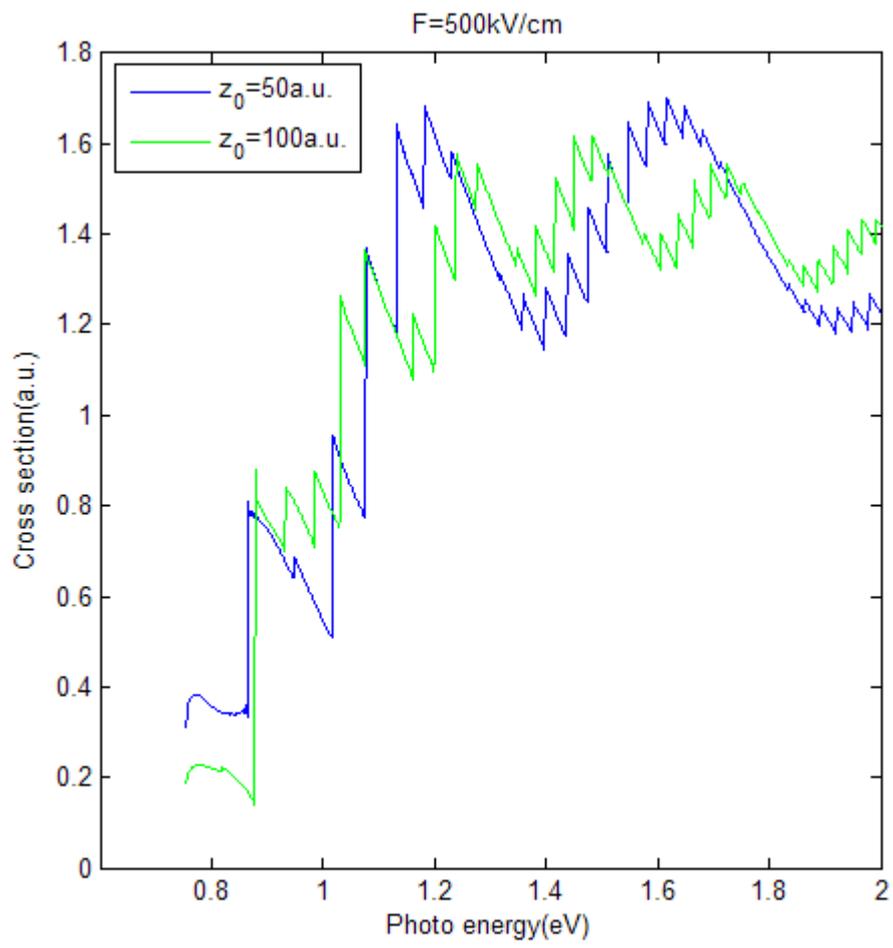

Fig.3

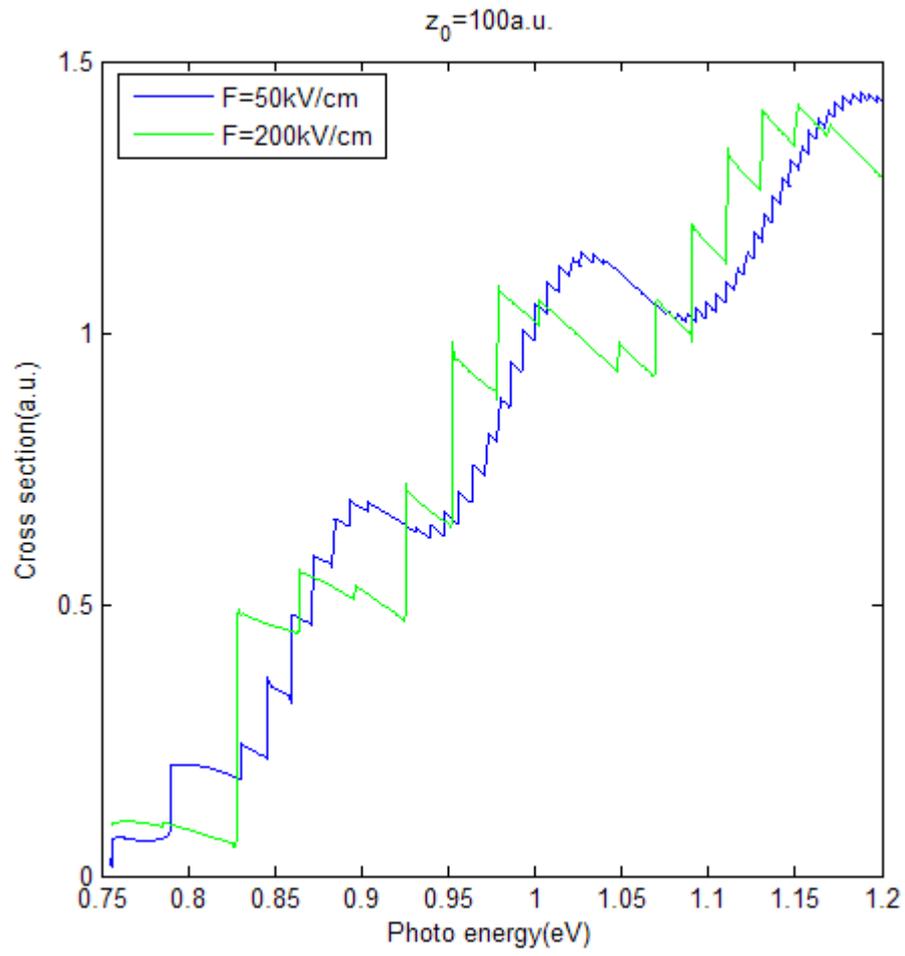

Fig.4

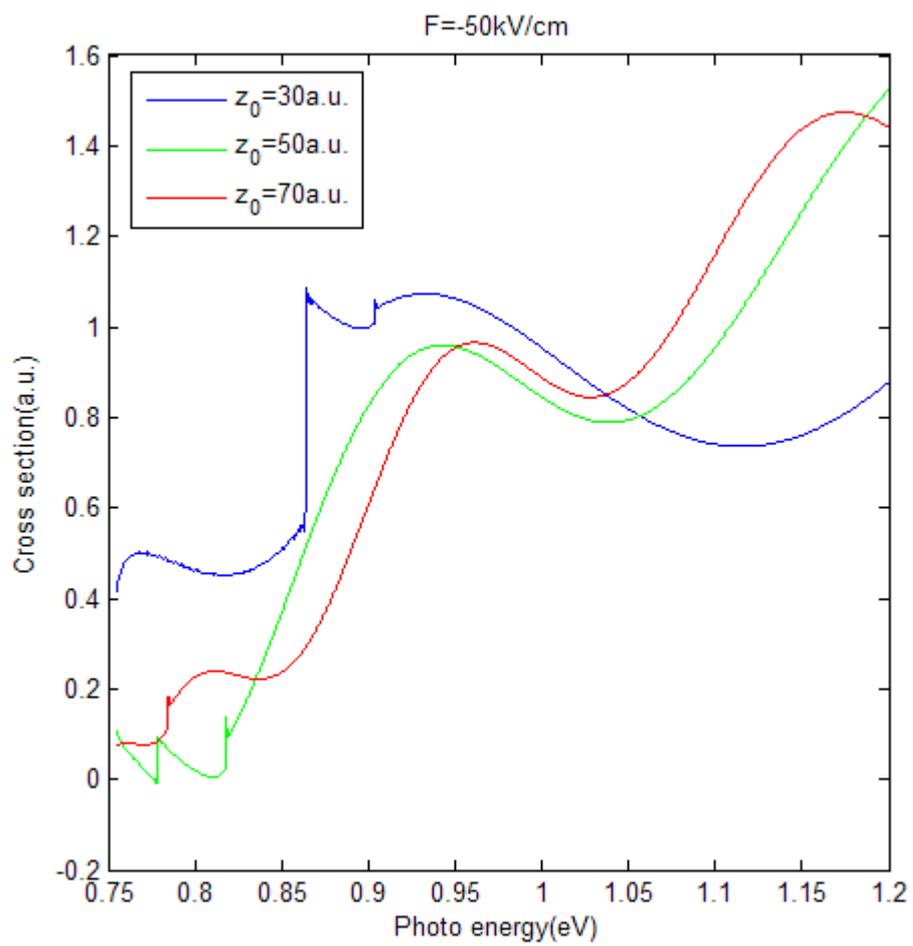

Fig.5

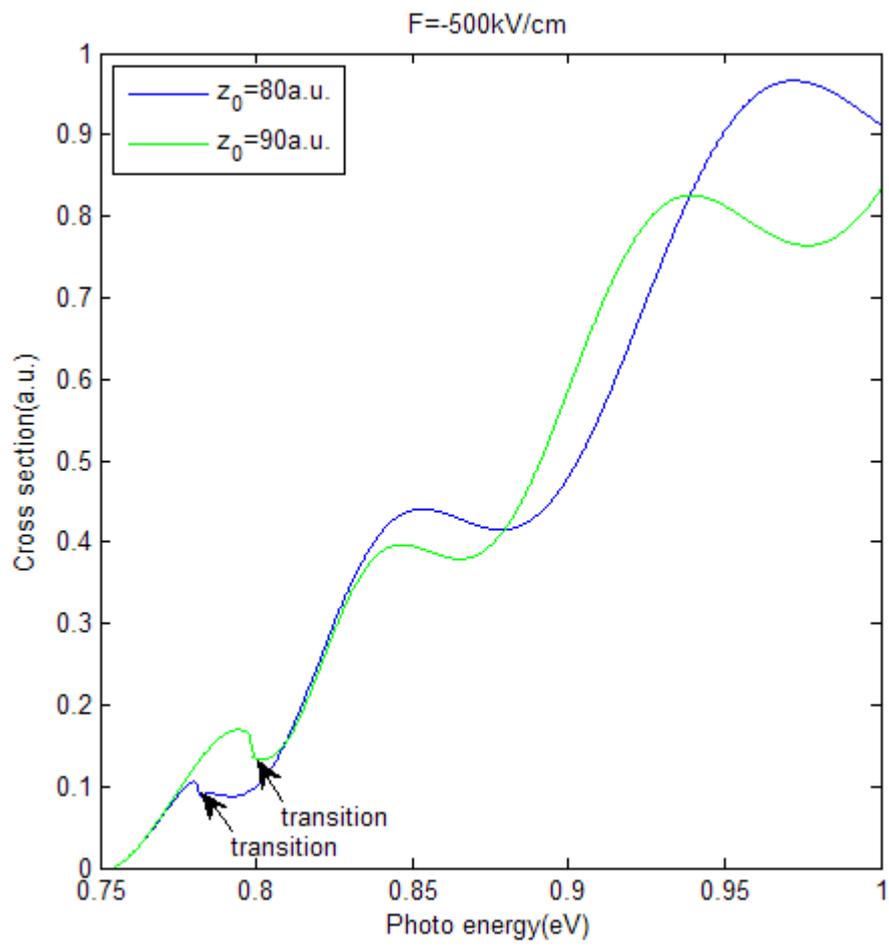

Fig.6